\pdfoutput=1

\documentclass{article}

\usepackage[final]{nips_2016}

\usepackage[utf8]{inputenc} 
\usepackage[T1]{fontenc}    
\usepackage{hyperref}       
\usepackage{url}            
\usepackage{booktabs}       
\usepackage{amsfonts}       
\usepackage{nicefrac}       
\usepackage{microtype}      

\usepackage{graphicx}
\usepackage{subfigure}
\usepackage{amsmath}

\hypersetup{hidelinks}

\title{Cardiac MRI Orientation Recognition and Standardization using Deep Neural Networks}

\author{
  Ruoxuan Zhen\\
  School of Data Science\\
  Fudan University, Shanghai\\
  \texttt{20307110408@fudan.edu.cn} \\
}

\begin{document}

\maketitle

\begin{abstract}
  Orientation recognition and standardization play a crucial role in the effectiveness of medical image processing tasks. Deep learning-based methods have proven highly advantageous in orientation recognition and prediction tasks. 
  In this paper, we address the challenge of imaging orientation in cardiac MRI and present a method that employs deep neural networks to categorize and standardize the orientation. 
  To cater to multiple sequences and modalities of MRI, we propose a transfer learning strategy, enabling adaptation of our model from a single modality to diverse modalities. 
  We conducted comprehensive experiments on CMR images from various modalities, including bSSFP, T2, and LGE. The validation accuracies achieved were 100.0\%, 100.0\%, and 99.4\%, confirming the robustness and effectiveness of our model. 
  Our source code and network models are available at https://github.com/rxzhen/MSCMR-orient
  \end{abstract}

\section{Introduction}

Cardiac Magnetic Resonance (CMR) images may exhibit variations in image orientations when recorded in DICOM format and stored in PACS systems. Recognizing and comprehending these differences are of crucial importance in deep neural network (DNN)-based image processing and computation, as DNN systems typically treat images merely as matrices or tensors, disregarding the imaging orientation and real-world coordinates.
This study aims to investigate CMR image orientation, with a focus on referencing human anatomy and a standardized real-world coordinate system. 
The goal is to develop an efficient method for recognizing and standardizing the orientation of CMR images. 
By achieving this goal, we can ensure consistency and enhance the accuracy of DNN-based image analysis in the context of cardiac MRI.

For CMR images, standardization of their orientations is a 
prerequisite for subsequent computing tasks utilizing DNN-based methodologies, 
such as image segmentation [4] and myocardial pathology analysis [1]. 
Deep learning methods have found widespread use 
in orientation recognition and prediction tasks. 
For instance, Wolterink et al. introduced an algorithm that 
employs a Convolutional Neural Network (CNN) to extract coronary 
artery centerlines in cardiac CT angiography (CCTA) images [5]. 
Building upon CMR orientation recognition, 
our work focuses on developing a method for standardizing and 
adjusting the image orientations.

\begin{figure}[h]
  \includegraphics[width=\textwidth]{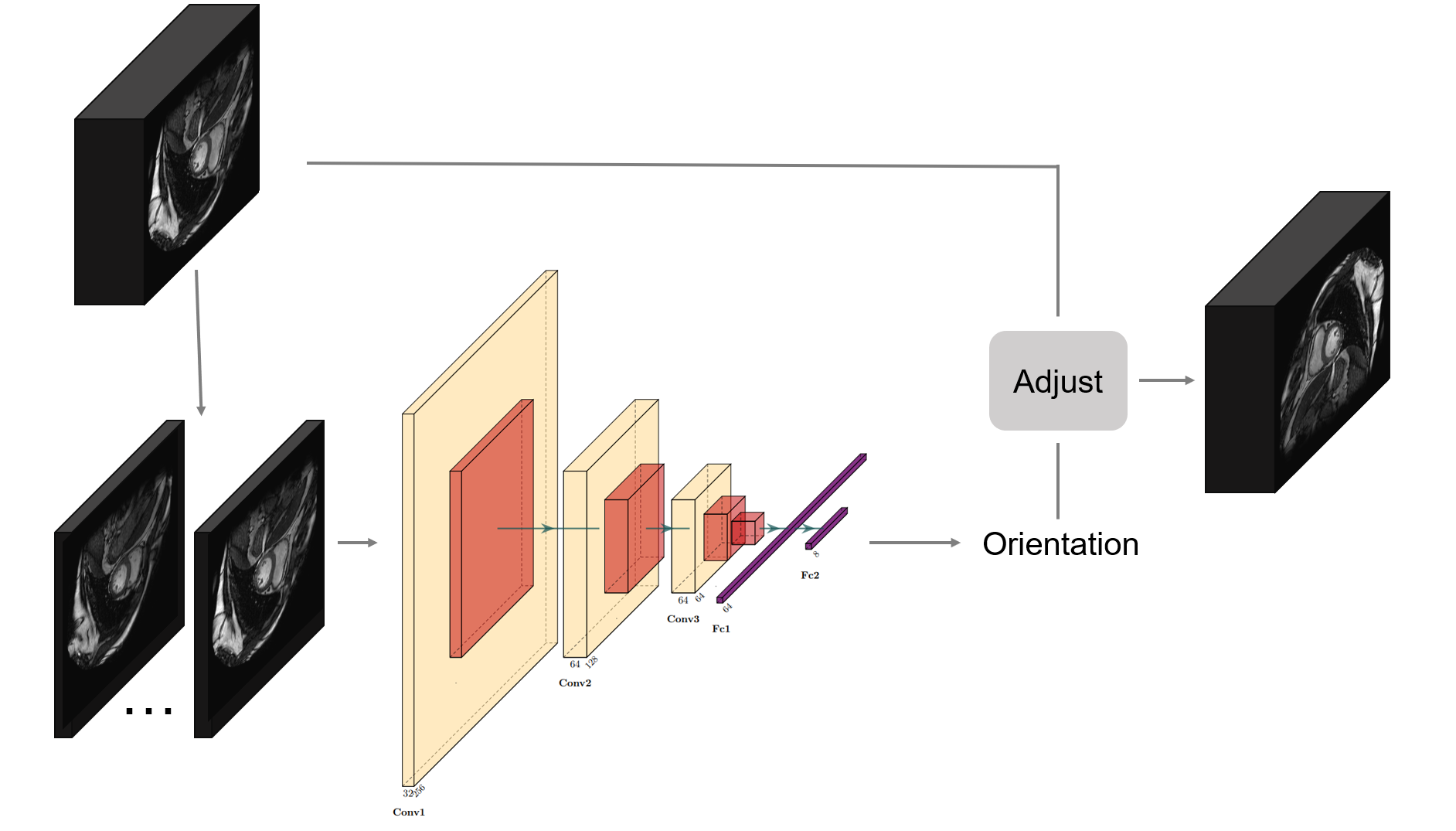}
  \caption{The pipeline of the proposed CMR orientation recognition and standardization method. 
  Initially, the image undergoes pre-processing. Subsequently, the image is input into a CNN to generate a orientation prediction. 
  Guided by this orientation prediction, the adjust tool can standardize the image, ensuring its alignment with the desired orientation.}
  \label{fig1}
\end{figure}  

This study aims to design a DNN-based approach for 
achieving orientation recognition and standardization 
across multiple CMR modalities. 
Figure~\ref{fig1} illustrates the pipeline of our proposed method. 
The key contributions of this work are summarized as follows:

\begin{itemize}
  \item[(1)] We propose a scheme to standardize the CMR image orientations and categorize them for classification purposes.
  \item[(2)] We present a DNN-based orientation recognition method tailored for CMR images and demonstrate its transferability to other modalities.
  \item[(3)] We develop a CMR image orientation adjustment tool embedded with a orientation recognition network. This tool greatly facilitates CMR image orientation recognition and standardization in clinical and medical image processing practices.
\end{itemize}

\section{Method}

In this section, we introduce our proposed method for orientation recognition and standardization. Our proposed framework is built on the categorization of CMR image orientations. We propose a DNN to recognize the orientation of CMR images and embed it into the CMR orientation adjust tool. 

\subsection{CMR Image Orientation Categorization}

Due to different data sources and scanning habits, the orientation of different cardiac magnetic resonance images may be different, and the orientation vector corresponding to the image itself may not correspond correctly. This may cause problems in tasks such as image segmentation or registration.
Taking a 2D image as an example, we set the orientation of an image as the initial image and set the four corners of the image to 
\framebox[1cm][c]{$\begin{array}{cc} 1 & 2 \\ 3 & 4 \end{array}$}, then the orientation of the 2D MR image may have the following 8 variations, which is listed in Table ~\ref{tab1}. 
For each image-label pair $(X_t,\ Y_t)$, we can flip $X_t,\ Y_t$ towards a picked orientation 
to get a new image-label pair. 
If we correctly recognize the orientation of an image, we can perform the reverse flip to standardize it.

\begin{table}[h]
\begin{center}
\begin{tabular}{c c c c }
\hline
Label &	Operation &	Image &	Correspondence of coordinates\\
\hline
0 &	Initial state & \framebox[1cm][c]{$\begin{array}{cc} 1 & 2 \\ 3 & 4 \end{array}$} & $\text{Target}[x,y,z]=\text{Source}[x,y,z]$\\
1 &	Horizontal flip &\framebox[1cm][c]{$\begin{array}{cc} 2 & 1 \\ 4 & 3 \end{array}$}& $\text{Target}[x,y,z]=\text{Source}[sx-x,y,z]$\\
2 &	Vertical flip &\framebox[1cm][c]{$\begin{array}{cc} 3 & 4 \\ 1 & 2 \end{array}$}& $\text{Target}[x,y,z]=\text{Source}[x,sy-y,z]$\\
3 &	Rotate $180^\circ$ clockwise &\framebox[1cm][c]{$\begin{array}{cc} 4 & 3 \\ 2 & 1 \end{array}$} & $\text{Target}[x,y,z]=\text{Source}[sx-x,sy-y,z]$\\
4 &	Flip along the main diagonal &\framebox[1cm][c]{$\begin{array}{cc} 1 & 3 \\ 2 & 4 \end{array}$}& $\text{Target}[x,y,z]=\text{Source}[y,x,z]$\\
5 &	Rotate $90^\circ$ clockwise &\framebox[1cm][c]{$\begin{array}{cc} 3 & 1 \\ 4 & 2 \end{array}$}& $\text{Target}[x,y,z]=\text{Source}[sx-y,x,z]$\\
6	& Rotate $270^\circ$ clockwise &\framebox[1cm][c]{$\begin{array}{cc} 2 & 4 \\ 1 & 3 \end{array}$}	& $\text{Target}[x,y,z]=\text{Source}[y,sy-x,z]$\\
7	& Flip along the secondary diagonal &\framebox[1cm][c]{$\begin{array}{cc} 4 & 2 \\ 3 & 1 \end{array}$} & $\text{Target}[x,y,z]=\text{Source}[sx-y,sy-x,z]$\\
\hline
\end{tabular}
\end{center}
\caption{Orientation Categorization of 2D CMR Images. Here, $sx$, $sy$ and $sz$ respectively denote the size of image in X-axis, Y-axis and Z-axis.}
\label{tab1}
\end{table}

\subsection{Deep Neural Network}

We employ a classical convolutional neural network for orientation recognition. It is a widely adopted approach in image classification tasks, adhering to the standard design pattern for CNNs.
The neural network architecture comprises 3 convolutional blocks, 
each housing a convolutional layer, batch normalization, 
ReLU activation, and max pooling. 
These blocks effectively capture features from the input images. Additionally, an average pooling layer and 2 fully connected layers, with 8 units in the output layer, complete the network, enabling orientation prediction.

To train the model effectively, we utilize the cross-entropy loss, which efficiently measures the discrepancy between the predicted orientation and the ground truth orientation label.

\subsection{Transfer Learning}

When adapting the proposed orientation recognition network to new datasets of different modalities, 
we employ a transfer learning approach to obtain the transferred model. 
Initially, we freeze the weights of the convolutional layers and fine-tune the fully connected layers. 
We repeat this process for the subsequent fine-tuning steps until the model converges. 
Afterwards, we unfreeze the weights of all layers and proceed to fine-tune the entire model.

\section{Experiment}

\subsection{Dataset}

We experiment with the MyoPS dataset [1,2], which provides the three-sequence CMR (LGE, T2, and bSSFP) from the 45 patients. 
We divide the CMR data of 45 patients into 
training and validation sets at the ratio of 80\% and 20\%. 

\subsection{Data Pre-processing}

For each CMR data, we initially apply 7 transformations according to Table 1 to ensure the dataset encompasses all 8 orientations. Subsequently, we slice the 3D CMR data into multiple 2D data instances.

Given an image-label pair $(X_t, Y_t)$, for each $X_t$, 
we identify the maximum gray value, denoted as $G$. 
Subsequently, three truncation operations are performed on $X_t$ 
using thresholds of $0.6G, 0.8G,$ and $G$ to generate ${X_1}_t, {X_2}_t,$ and ${X_3}_t$, respectively. In this operation, pixels with gray values higher than the threshold are mapped to the threshold gray value. The utilization of different thresholds allows us to capture image characteristics under various gray value window widths, mitigating the impact of extreme gray values. Additionally, grayscale histogram equalization is applied to ${X_1}_t, {X_2}_t,$ and ${X_3}_t$, resulting in ${X'_1}_t, {X'_2}_t,$ and ${X'_3}_t$. Finally, we concatenate these three 2D images into a 3-channel image $[{X'_1}_t, {X'_2}_t, {X'_3}_t]$, which serves as the input to our proposed DNN.

We perform data augmentation by randomly rotating the image slightly and applying random crops and resizing. These approaches introduce variability in the orientation of the images, which aids in improving model generalization and enhances robustness to varying image sizes and aspect ratios.

\subsection{Results}

\begin{figure}[h]
  \centering
  \subfigure[bSSFP]{\includegraphics[height=3cm]{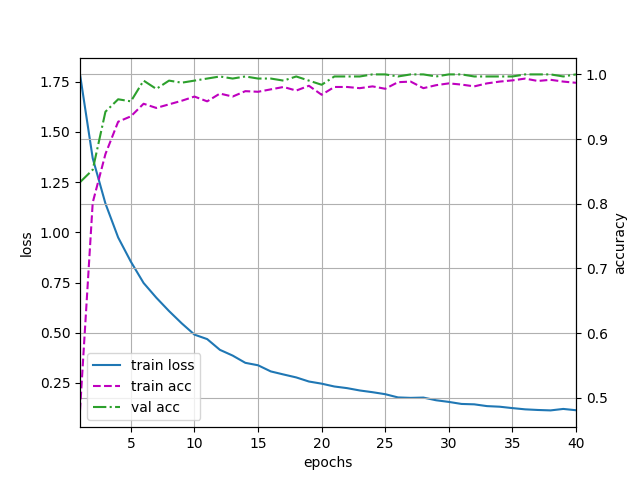}}
  \subfigure[T2]{\includegraphics[height=3cm]{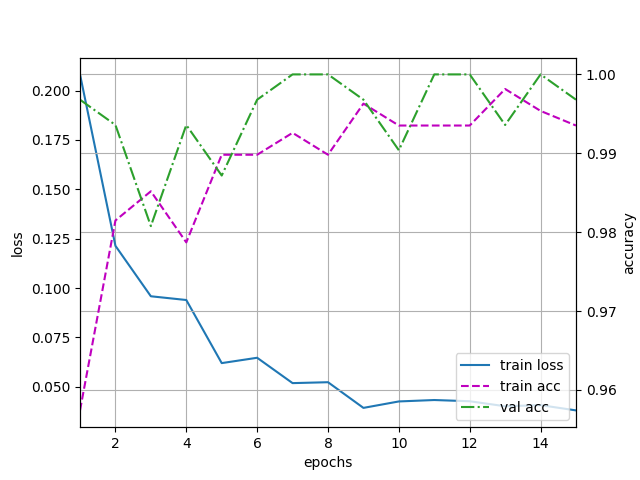}}
  \subfigure[LGE]{\includegraphics[height=3cm]{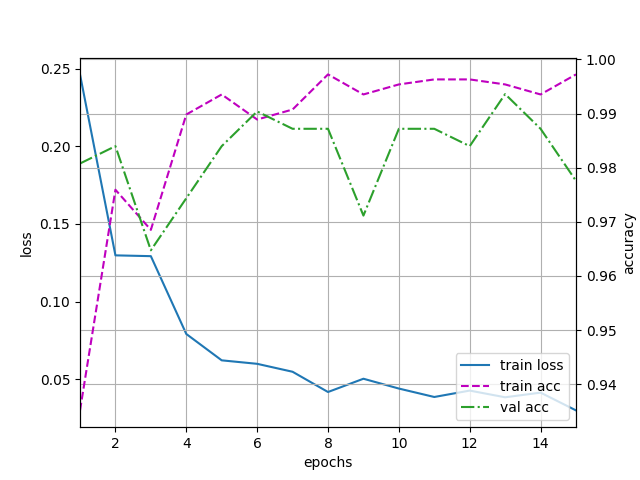}}
  \caption{Training process. On the bSSFP modality, we trained for 40 epochs, and subsequently fine-tuned on other modalities for 15 epochs, achieving remarkably high accuracy.}
  \label{fig2}
\end{figure}

We initially train the model on the bSSFP modality and subsequently fine-tune it on the T2 and LGE modalities. The training process is depicted in Figure~\ref{fig2}. The average accuracy on the dataset is presented in Table~\ref{tab2}.
The results highlight the model’s ability to transfer learning to other modalities,
showcasing a remarkable level of accuracy.

\begin{table}[h]
\begin{center}
\begin{tabular}{ccc}
\hline
Modality &	Train accuracy &	Validation accuracy\\
\hline
bSSFP & 99.4\% & 100.0\% \\
LGE & 99.5\% & 99.4\% \\
T2 &  99.5\% & 100.0\% \\
\hline
\end{tabular}
\end{center}
\caption{Orientation recognition accuracy of 2D MS-CMR. For each 3D CMR data, by selecting the most frequent orientation prediction from the predicted orientations of its 2D slices, we can achieve a 100\% accuracy on both the training and validation sets for all modalities.}
\label{tab2}
\end{table}

\section{Conclusion}

We have introduced a DNN-based method for multi-sequence MRI images. 
The experimental results validate the effectiveness of the 
orientation recognition network in accurately classifying the orientation of multi-sequence CMR images. 
Our future research  will focus on expanding the categorization of the CMR image
orientation and refining the classification network to further enhance the classification accuracy.

\section*{References}

\medskip

\small

[1] Xiahai Zhuang: Multivariate mixture model for myocardial segmentation combining multi-source images. IEEE Transactions on Pattern Analysis and Machine Intelligence 41(12), 2933–2946, 2019 

[2] Junyi Qiu, Lei Li, Sihan Wang, Ke Zhang, Yinyin Chen, Shan Yang, Xiahai Zhuang. MyoPS-Net: Myocardial Pathology Segmentation with Flexible Combination of Multi-Sequence CMR Images. Medical Image Analysis 84, 102694, 2023 

[3] Ke Zhang and Xiahai Zhuang: Recognition and Standardization of Cardiac MRI Orientation via Multi-tasking Learning and Deep Neural Networks. MyoPS 2020, LNCS 12554, 167–176, Springer Nature, 2020
	https://github.com/BWGZK/Orientation-Adjust-Tool

[4] Zhuang X, Li L, Payer C, et al. Evaluation of algorithms for multi-modality whole heart segmentation: an open-access grand challenge[J]. Medical image analysis, 2019, 58: 101537.

[5] Wolterink J M, van Hamersvelt R W, Viergever M A, et al. Coronary artery centerline extraction in cardiac CT angiography using a CNN-based orientation classifier[J]. Medical image analysis, 2019, 51: 46-60.

\end{document}